\documentclass[floats,floatfix,showpacs,amssymb,prd,superscriptaddress,nofootinbib,twocolumn,aps]{revtex4}

\usepackage{graphicx, epsfig, amssymb} 
\usepackage{amsmath, amsfonts}
\usepackage{bm} 

\usepackage[linktocpage]{hyperref}
\usepackage[caption=false]{subfig}
\usepackage[usenames]{color}

\def\be{\begin{equation}}
\def\ee{\end{equation}}

\newcommand{\bea}{\begin{eqnarray}}
\newcommand{\eea}{\end{eqnarray}}
\newcommand{\ben}{\begin{enumerate}}
\newcommand{\een}{\end{enumerate}}
\newcommand{\bi}{\begin{itemize}}
\newcommand{\ei}{\end{itemize}}

\newcommand{\nn}{\nonumber}

\def\ga{\mathrel{\raise.3ex\hbox{$>$\kern-.75em\lower1ex\hbox{$\sim$}}}}
\def\la{\mathrel{\raise.3ex\hbox{$<$\kern-.75em\lower1ex\hbox{$\sim$}}}}

\def\l{\left}
\def\r{\right}
\def\be{\begin{equation}}
\def\ee{\end{equation}}

\def\I_M{{I_{\scriptscriptstyle M\times M}}}

\def\be{\begin{equation}}
\def\ee{\end{equation}}
\def\bea{\begin{eqnarray}}
\def\eea{\end{eqnarray}}
\newcommand{\beq}{\begin{eqnarray}}
\newcommand{\eeq}{\end{eqnarray}}

\newcommand{\beqal}{\begin{eqnarray}\label}
\newcommand{\beqa}{\begin{eqnarray}}
\newcommand{\eeqa}{\end{eqnarray}}

\begin{document}\title{\large Slowly rotating black holes in alternative theories of gravity}

\author{Paolo Pani}
\affiliation{CENTRA, Departamento de F\'{\i}sica, 
Instituto Superior T\'ecnico, Universidade T\'ecnica de Lisboa - UTL,
Av.~Rovisco Pais 1, 1049 Lisboa, Portugal.}

\author{Caio F. B. Macedo}
\affiliation{Faculdade de F\'{\i}sica, Universidade Federal do Par\'a, 66075-110, Bel\'em, Par\'a, Brazil.}

\author{Lu\'{\i}s C. B. Crispino}
\affiliation{Faculdade de F\'{\i}sica, Universidade Federal do Par\'a, 66075-110, Bel\'em, Par\'a, Brazil.}

\author{Vitor Cardoso}
\affiliation{CENTRA, Departamento de F\'{\i}sica, 
Instituto Superior T\'ecnico, Universidade T\'ecnica de Lisboa - UTL, Av.~Rovisco Pais 1, 1049 Lisboa, Portugal.}
\affiliation{Department of Physics and Astronomy, The University of Mississippi, University, MS 38677, USA.}

\begin{abstract}
We present, in closed analytic form, a general stationary, slowly rotating black hole, which is solution to a large class of alternative theories of gravity in four dimensions. In these theories, the Einstein-Hilbert action is supplemented by all possible quadratic, algebraic curvature invariants coupled to a scalar field. The solution is found as a deformation of the Schwarzschild metric in General Relativity.
We explicitly derive the changes to the orbital frequency at the innermost stable circular orbit and at the light ring in closed form. These results could be useful when comparing General Relativity against alternative theories by (say) measurements of X-ray emission in accretion disks, or by stellar motion around supermassive black holes. When gravitational-wave astronomy comes into force, strong constraints on the coupling parameters can in principle be made.
\end{abstract}


\maketitle
\date{today}

\section{Introduction}
General Relativity (GR) is an elegant theory which agrees with all observations at Solar System scale and beyond~\cite{Will:2005va,Everitt:2011hp}; however its nonlinear, strong-field structure still remains elusive and difficult to test~\cite{lrr-2008-9}. This, together with some long-standing problems in Einstein theory
(like the presence of singularities, difficulties in explaining the accelerated universe and galaxy rotation curves, etc),  
has motivated the study of viable alternative theories of gravity. These theories, also known as modified theories of gravity, aim to reproduce GR in the weak-field regime, but they can differ substantially from it in the strong curvature regime, where nonlinear effects become dominant. 
In order to pass current experiments, alternative theories should have the same post-Newtonian expansion as GR, at least to lowest order. However, large deviations are possible in relativistic systems: black holes (BHs), neutron stars, and cosmological models.

BHs are natural candidates to investigate strong curvature corrections to GR. In the next decade, gravitational-wave detectors~\cite{Berti:2004bd} and high-frequency very long baseline interferometry (VLBI)~\cite{Doeleman:2008xq} may provide direct observations of these objects and of their nonlinear structure, completing the wealth of information from current electromagnetic observations~\cite{lrr-2008-9}. The geometric structure of BHs encodes information about the underlying theory of gravity. Within GR, no-hair theorems (see Ref.~\cite{Bekenstein:1995un} and references therein) guarantee that stationary BHs are described by the Kerr solution and this assumption enters most of the calculations, including gravitational-wave emission, gravitational lensing and properties of the accretion disks. 
However, when corrections to GR are considered, BHs can support non-trivial hairs~\cite{Kanti:1995vq} and new classes of solutions may exist. Hence, it is important to derive deformations to the Kerr metric~\cite{Collins:2004ex,Vigeland:2011ji,Johannsen:2011dh} arising from alternative theories of gravity and to predict astrophysical observables within a more general, bias-independent framework. 

Previous studies on BH solutions in alternative theories of gravity suffer from two majors limitations.
First, given the plethora of alternative theories that have been recently proposed, most of the approaches have focused on a case-by-case analysis (with the notable exceptions of Refs.~\cite{Vigeland:2011ji,Johannsen:2011dh}). Secondly, motivated and well-behaved corrections to GR are usually involved, so that BHs must be constructed numerically. In particular, rotating solutions are extremely challenging to find in closed form and the Kerr metric is usually regarded as unique in this context. Thus, analytical solutions describing rotating BHs in a broad class of alternative theories, as the one we present here, are of utmost importance.
 
In this work, generalizing previous studies on static BHs~\cite{Yunes:2011we}, we derive the metric of slowly rotating BHs arising as solutions of a large class of alternative theories of gravity, in which the Einstein-Hilbert action is supplemented by all quadratic, algebraic curvature terms coupled to a scalar field.
Rotating BH solutions are relevant for several reasons. Astrophysical BHs are likely to be (rapidly) spinning, due to accretion effects. Thus, any realistic computation (for example the properties of accretion disks) must take rotation into account. 
Furthermore, the imprints of possible strong curvature corrections are expected to be stronger for those processes taking place close to near-extremal rotating BHs, for which the curvature is larger. For example, the Kretschmann invariant, ${\cal K}=R_{abcd}R^{abcd}$, on the equatorial event horizon of a Kerr BH of mass $M$ and angular momentum $J=aM$ in Boyer-Lindquist coordinates reads 
${\cal K}={48M^2}{\left[M+\sqrt{M^2-a^2}\right]^{-6}}$, where here and in the rest of the paper we use $G=c=1$ units. For a Schwarzschild BH ($a=0$), ${\cal K}M^4=3/4$. However, for extremal Kerr BHs ($a=M$) this scalar invariant is $\sim60$ times larger, ${\cal K}M^4=48$.


\section{Gravity with quadratic curvature corrections}
We consider a class of alternative theories of gravity in four dimensions obtained by including all quadratic, algebraic curvature invariants, generically coupled to a single scalar field~\cite{Yunes:2011we}. The action of this theory reads
%
\begin{eqnarray}
 S&&=\frac{1}{16\pi}\int\sqrt{-g} d^4x \left[R-2\nabla_a\phi\nabla^a\phi-V(\phi)+f_1(\phi)R^2\right.\nn\\
&&\left.+f_2(\phi) R_{ab}R^{ab}+f_3(\phi) R_{abcd}R^{abcd}+f_4(\phi)R_{abcd}{}^*R^{abcd}\right]\nn\\
&&+S_\text{mat}\left[\gamma(\phi)g_{\mu\nu},\Psi_\text{mat}\right]\,,\label{action}
\end{eqnarray}
%
where, in the matter action $S_\text{mat}$, we have generically included a non-minimal coupling, which naturally arises in some string theories defined in the Einstein frame. In the following, we neglect the scalar self-potential $V(\phi)$. Its inclusion, along with theories in asymptotically non-flat spacetimes, is a natural extension of the present work.

When $f_1=\alpha e^{-2\phi}$, $f_2=-4f_1$ and $f_3=f_1$, the theory reduces to the bosonic sector of heterotic string theory and the quadratic corrections reduce to the Gauss-Bonnet invariant. In that case matter is non-minimally coupled to gravity, $\gamma(\phi)=e^{\phi}$. Static BH solutions in Gauss-Bonnet gravity were found analytically in the small coupling limit~\cite{Campbell:1991kz,Mignemi:1992pm} and numerically for general coupling~\cite{Kanti:1995vq} (see also Ref.~\cite{Torii:1996yi}). Stationary BHs with Gauss-Bonnet corrections were considered numerically in Ref.~\cite{Pani:2009wy} for slow rotations, whereas their highly spinning counterpart was recently constructed in Ref.~\cite{Kleihaus:2011tg}.
Furthermore, when $f_1=f_2=f_3=0$ and $f_4=\alpha_4\phi$, the above theory reduces to Chern-Simons gravity~\cite{Alexander:2009tp} and slowly rotating BHs in this theory where obtained in Refs.~\cite{Yunes:2009hc,Konno:2009kg}. The field equations arising from Eq.~\eqref{action} are explicitly given in Ref.~\cite{Yunes:2011we}, where analytical, static BH solutions were also obtained in the small coupling limit.
Here we generalize previous studies by constructing slowly rotating BHs in the general theory~\eqref{action}.

The theory~\eqref{action} has to be considered as an effective action, obtained as a truncation from a more general theory. For example in the low-energy expansion of some string theories, the Gauss-Bonnet and Chern-Simons terms arise as second order corrections in curvature. The Einstein-Hilbert term is considered as the first order term in a (possibly infinite) series expansion containing all possible curvature corrections. In this sense, GR may be only accurate up to ${\cal O}\left(\alpha R^2\right)$ and second order corrections may be important when dealing with nonlinear, relativistic solutions. For the same reason, we  work in a perturbative regime in which possible higher order terms in~\eqref{action} can be safely neglected. 
We consider the weak-field expansion of the coupling functions
\begin{equation}
 f_i(\phi)=\eta_i+\alpha_i \phi+{\cal O}(\phi^2)\,,\qquad i=1,2,3,4\nn\\
\end{equation}
where $\eta_i$ and $\alpha_i$ are dimensionful coupling constants. When the coupling functions are constant, i.e. $\alpha_i=0$, the theories above are usually labeled ``non-dynamical'' and they admit all vacuum GR solutions~\cite{Yunes:2011we}. As a result, for small scalar fields the background solutions do not depend on $\eta_i$. Although non-dynamical theories would have a different linear response, for example a different gravitational-wave emission~\cite{Yunes:2007ss,Molina:2010fb}, here we are interested in modified \emph{background} solutions and we then focus on dynamical couplings. Remarkably, in the small coupling limit, the dynamical theory only depends on four couplings, $\alpha_i$, regardless the coupling functions $f_i(\phi)$.
\section{Slowly rotating Black holes}
We consider the following metric ansatz for the stationary, slowly rotating limit, 
\begin{eqnarray}
 ds^2=&&-f(r,\theta)dt^2+g(r,\theta)^{-1}dr^2-2\omega(r)\sin^2\theta dtd\varphi+\nn\\
&&+r^2\Theta(r,\theta)d\theta^2+r^2\sin^2\theta\Phi(r,\theta) d\varphi^2 \label{metric}
\end{eqnarray}
together with the scalar field $\phi=\phi(r,\theta)$.
In Appendix~\ref{app:derivation} we solve the field equations~\cite{Yunes:2011we} order by order in a perturbative scheme for slow rotations and small couplings. Here, we simply report the final result. We obtain that the slowly rotating BH metric functions read
\begin{widetext}
 \begin{eqnarray}
f(r,\theta)&=& 1-\frac{2 M}{r}+\frac{\alpha_3^2}{4}\left[-\frac{49}{40 M^3 r}+\frac{1}{3 M r^3}+\frac{26}{3 r^4}+\frac{22 M}{5 r^5}+\frac{32 M^2}{5 r^6}-\frac{80 M^3}{3 r^7}\right]+a^2\frac{2M\cos^2\theta}{r^3}\,,\label{g00}\\ 
g(r,\theta)&=& 1-\frac{2 M}{r}+\frac{\alpha_3^2}{4}\left[-\frac{49}{40 M^3 r}+\frac{1}{M^2 r^2}+\frac{1}{M r^3}+\frac{52}{3 r^4}+\frac{2 M}{r^5}+\frac{16 M^2}{5 r^6}-\frac{368 M^3}{3 r^7}\right]+a^2\frac{r-(r-2M)\cos^2\theta}{r^3} \,,\\ 
\omega(r)&=&\frac{2 a M}{r}-\frac{a\alpha_3^2}{4}\left[\frac{3}{5 M r^3}+\frac{28}{3 r^4}+\frac{6 M}{r^5}+\frac{48  M^2}{5 r^6}-\frac{80 M^3}{3 r^7}\right]-a\alpha_4^2\frac{5}{2}\left[\frac{1}{r^4}+\frac{12 M}{7r^5}+\frac{27 M^2}{10r^6}\right] \,, \label{crosscomponent}\\
\Theta(r,\theta)&=& 1+\frac{\cos^2\theta}{r^2}a^2\,,\qquad \Phi(r,\theta)=1+\frac{r+2M\sin^2\theta}{r^3}a^2 \,,
\end{eqnarray}
whereas the scalar field reads
\begin{eqnarray}
\phi(r,\theta)&=&{\alpha_3}\left[\frac{1}{2M r}+\frac{1}{2r^2}+\frac{2 M}{3 r^3}\right]+a\alpha_4 \frac{5\cos\theta}{8M}\left[\frac{1}{r^2}+\frac{2 M}{r^3}+\frac{18 M^2}{5r^4}\right]\nn\\
&&-\frac{\alpha_3 a^2}{2}\left[\frac{1}{10r^4}+\frac{1}{5Mr^3}+\frac{1}{4M^2r^2}+\frac{1}{4M^3r}+\cos^2\theta \left(\frac{48M}{5r^5}+\frac{21}{5r^4}+\frac{7}{5Mr^3}\right)\right] \,,\label{scalar}
\end{eqnarray}
\end{widetext}
%
where the novel terms are those proportional to $a\alpha_3^2$ and to $a^2\alpha_3$ in Eq.~\eqref{crosscomponent} and Eq.~\eqref{scalar}, respectively. Interestingly, these terms are the dominant corrections at large distances, because they scale with a lower power of $r$ than those proportional to $\alpha_4$.
As explained in the appendix, the metric is found by requiring asymptotic flatness and regularity for $r>0$. The curvature invariants are regular in the exterior spacetime.
The BH mass and angular momentum can be read off from the $1/r$ coefficients in Eqs.~\eqref{g00} and~\eqref{crosscomponent}. The angular momentum reads $J=aM$, whereas the physical mass of the BH is ${\cal M}=M(1+{49}\alpha_3^2/{(320M^4)})$~\cite{Yunes:2011we}.
The above solution is accurate up to order ${\cal O}(a^2/M^2,\alpha_i^2/M^4,a\alpha_i^2/M^5)$ in the metric and up to order ${\cal O}(a^2/M^2,\alpha_i^2/M^4,a\alpha_i^2/M^5,a^2\alpha_i/M^3)$ in the scalar field. At this order, the angular metric functions $\Theta$ and $\Phi$ are simply given by the slowly rotating Kerr solution. For $a=0$, the slowly rotating BH correctly reduces to the static one setting, in the notation of Ref.~\cite{Yunes:2011we}, $\alpha_i=\alpha_{i}/(16\pi)$, $\beta=1/(4\pi)$, $\kappa=1/(16\pi)$. Furthermore, for $\alpha_3=0$, it reduces to the slowly rotating Chern-Simons BH~\cite{Yunes:2009hc}.
Slowly rotating black holes in Einstein-Dilaton-Gauss-Bonnet gravity can obtained from our solution setting $\alpha_4=0$ and their exact metric is presented here for the first time. We have compared the analytical results with numerical solutions~\cite{Pani:2009wy}, finding very good agreement.
Interestingly, this solution only depends on the couplings $\alpha_3$ and $\alpha_4$, since the terms proportional to $\alpha_1$ and $\alpha_2$ do not contribute to this order. Moreover, the corrections to the scalar field arising from $\alpha_3$ and $\alpha_4$ enter at different order in $a$: the Kretschmann correction only introduces even powers of $a$, while the Chern-Simons term only introduces odd powers~\cite{Yunes:2011we}. Nevertheless, both corrections affect the gravitomagnetic part of the metric, for example giving a modified frame-dragging effect. Finally, the corrections proportional to $\alpha_3$ scale with a lower power of $r$ than  those proportional to $\alpha_4$. Hence, they are expected to be dominant at large distances.

\section{Geodesic structure}
Many interesting and potentially observable effects around astrophysical BHs ultimately depend on how particles move in the region few Schwarzschild radii away from the event horizon. For example, the inner properties of the accretion disk are strongly affected by the location of the innermost stable circular orbit (ISCO) and, in turn, by the geodesic structure of the underlying spacetime. 
Most of the computations assume that the spacetime is described by a Kerr BH. However, deformed solutions arising in alternative theories would also affect particle motion, with potentially observable consequences. 
In the modified theories considered here, test-particles follow spacetime geodesics. This follows from the conservation of the stress-energy tensor, $\nabla^\mu T_{\mu\nu}=0$, which is guaranteed by the diffeomorphism invariant action~\eqref{action}. In many situations the geodesic motion of massive and massless particles is enough to fully describe many effects of astrophysical interest.

We consider the following matter action for a point-like particle
\be
S_\text{mat}=-m\int dt\,\sqrt{-\gamma(\phi)g_{\mu\nu}\dot x^\mu\dot x^\nu}\,,\label{eq:mataction}
\ee
where $m$ is the mass of the particle and $\gamma(\phi)$ is a possible coupling function between the matter and the scalar field. For low-energy modifications from heterotic string theory, $\gamma=e^{\phi}$.
In the small field limit, we may write
\begin{equation}
 \gamma(\phi)=1+4b\phi+{\cal O}(\phi^2)\,,
\end{equation}
where $b=0$ for minimal coupling and $b=1/4$ in heterotic string theory. We focus on equatorial motion ($\theta=\pi/2$, $\dot\theta\equiv0$). The radial geodesic motion on the equatorial plane can be derived from the equation
\begin{equation}
 \dot{r}^2=V(r)= \frac{g}{\gamma^2}\left(\frac{h E^2-f L^2+2j E L}{j^2+fh}-\delta \gamma\right)\,,
\end{equation}
where $j=-\omega(r)$, $h=\Phi(r)r^2$, and $\delta=0,1$ for massless and massive particles, respectively. Here $E$ and $L$ are the energy per unit of mass and the angular momentum per unit of mass of the orbiting particle, respectively. 
For circular orbits at $r=r_c$, the corresponding values of $E$ and $L$ can be found by imposing 
$V(r_c,E_c,L_c)=0=V'(r_c,E_c,L_c)$
and, for $\delta=1$, the ISCO location is defined through $V''(r_\text{ISCO},E_c,L_c)=0$.
Finally, the frequency at the ISCO reads
\begin{equation}
\Omega_{\text{ISCO}}=\l.\frac{\dot{\varphi}}{\dot{t}}\r|_{r_\text{ISCO}}=\frac{f-j E_c/L_c}{h E_c/L_c+j}\,.
\end{equation}

In line with our approximation scheme, we expand the geodesic quantities around their Schwarzschild value, i.e. 
\begin{eqnarray}
 X&&=X^{(0)}+X^{(1)}a+X^{(2)}a^2+X^{(3)}\alpha_3+X^{(4)}\alpha_3^2\nn\\
&&+X^{(5)}a\alpha_3+X^{(6)}a\alpha_3^2+X^{(7)}a^2\alpha_3+X^{(8)}a\alpha_4^2\,, \label{circular}
\end{eqnarray}
where $X$ schematically denotes $r_c$, $E_c$ and $L_c$. In general, the coupling $b$ introduces lower order contributions, like those proportional to $\alpha_3$. This is due to the lower order dependence of the scalar field in Eq.~\eqref{scalar}. For the same reason, such corrections do not arise for terms proportional to $\alpha_4$, since the odd-parity correction to the scalar field vanishes on the equatorial plane. Substituting the expansion~\eqref{circular} and solving order by order, we obtain the following ISCO location and the frequency at the ISCO, normalized by the physical mass $\mathcal{M}$, 
\bea
\frac{r_\text{ISCO}}{{\cal M}}&&=6-4 \sqrt{\frac{2}{3}} \frac{a}{M}-\frac{7 a^2}{18 M^2}+\frac{16}{9}\frac{b\alpha_3}{M^2}-\frac{17}{27}\sqrt{\frac{2}{3}}\frac{ba\alpha_3}{M^3}\nn\\
&&-\left(\frac{16297}{38880}-\frac{22267 a}{17496 \sqrt{6} M}\right) \frac{\alpha_3^2}{M^4}+\frac{77 a}{216 \sqrt{6} M^5}\alpha_4^2,\nn\\
%
{\cal M}\Omega_\text{ISCO}&&=\frac{1}{6 \sqrt{6}}+\frac{11 a}{216 M}+\frac{59 a^2}{648 \sqrt{6} M^2}-\frac{12113 a}{5225472 M^5}\alpha_4^2\nn\\
&&-\frac{29}{216 \sqrt{6}}\frac{b\alpha_3}{M^2}-\frac{169}{3888}\frac{ba\alpha_3}{M^3}\nn\\
&&+\left(\frac{32159}{2099520 \sqrt{6}}-\frac{49981 a}{75582720 M}\right) \frac{\alpha_3^2}{M^4}\,,
\eea
where we have kept only dominant terms in $b$ and we are considering corotating orbits only. Counter-rotating orbits can be simply obtained by inverting the sign of $a$.
The behavior of the ISCO frequency depends on several couplings. For $b=0$, the dominant correction is ${\cal O}(\alpha_3^2)$ and contribute to increase the frequency. The first corrections proportional to the BH spin are ${\cal O}(a\alpha_3^2)$ and ${\cal O}(a\alpha_4^2)$ and they contribute to lower the frequency. However, when a non-minimal coupling is turned on, its effect is dominant~\cite{Pani:2009wy}. The ISCO frequency gets \emph{negative} ${\cal O}(b\alpha_3)$ corrections. Since this is the dominant effect, a decreasing of the ISCO frequency could be seen as a general signature of non-minimal couplings, regardless the relative strength of $a$, $\alpha_3$ and $\alpha_4$. 

The same procedure can be applied to null geodesics, which are the trajectories of massless particles. In this case, it is easy to show that the result does not depend on the coupling $\gamma$. We get
 \begin{eqnarray}
 \frac{r_{\rm null}}{{\cal M}}&&=3-\frac{2 a}{\sqrt{3} M}-\frac{2 a^2}{9 M^2}+\frac{31}{81 \sqrt{3}}\frac{a\alpha_4^2}{M^5}\nn\\
&&-\left(\frac{961}{3240}-\frac{33667 a}{174960 \sqrt{3} M}\right) \frac{\alpha_3^2}{M^4}\,,\\
{\cal M}\Omega_{\rm null}&&=\frac{1}{3 \sqrt{3}}+\frac{2 a}{27 M}+\frac{11 a^2}{162 \sqrt{3} M^2}-\frac{131}{20412} \frac{a\alpha_4^2}{M^5}\nn\\
&&+\left(\frac{4397}{262440 \sqrt{3}}+\frac{24779 a}{4723920 M}\right) \frac{\alpha_3^2}{M^4} \,,
\end{eqnarray}
where $\Omega_{\rm null}=L_{\rm null}/E_{\rm null}$ is the light-ring frequency, and it is related to the real part of the ringdown frequency of the BH in the eikonal limit~\cite{Cardoso:2008bp}.
The dominant correction is ${\cal O}(\alpha_3^2)$ and contributes to increase the frequency, whereas the ${\cal O}(a\alpha_3^2)$ and ${\cal O}(a\alpha_4^2)$ corrections have an opposite relative sign.
\section{Discussion and Conclusion}
We have found slowly rotating BHs, solutions of a class of alternative theories as general as the action~\eqref{action}. This theory supplements GR by all quadratic, algebraic curvature terms coupled to a scalar field.
Our solution is presented in closed form up to some order in the angular momentum and in the coupling parameters. To the same order, we discussed the most relevant properties of the equatorial geodesic motion, giving the ISCO and light-ring frequencies.

With the analytical solution at hand, several extensions of the present work are possible. 
The properties of the (modified) accretion disk can be used to constrain the parameters of the theory~\cite{lrr-2008-9}. 
Furthermore, the study of the geodesic structure can be generalized to include non-equatorial orbits and an analysis similar to Ref.~\cite{Amarilla:2010zq} can be performed. 
Another interesting issue is the linear response of the slowly rotating BH. Strong curvature corrections to GR affect the linear stability analysis~\cite{Pani:2009wy} and the gravitational-wave emission~\cite{Yunes:2009ke}.

In addition, several extensions of the present solution are conceivable. First of all, going further in the approximation scheme, up to order ${a^2\alpha_i^2}$, corrections to the event horizon location and to the ergoregion would appear. This can have a profound impact on the stability of these solutions.
Furthermore, highly spinning BHs are phenomenologically more relevant and larger deviations from the Kerr metric may be expected. However, they have to be constructed numerically~\cite{Kleihaus:2011tg} on the basis of a case-by-case analysis. In this case, our analytical solution can be useful; for example it can be used as an initial profile to start numerical relaxation methods, or to check numerical solutions.

We report here that the slowly rotating metric we found can be mapped into the bumpy BH formalism along the same line discussed in Ref.~\cite{Vigeland:2011ji}, although the mapping is non-trivial. On the other hand, this solution does not belong to the class of deformed Kerr BHs proposed in Ref.~\cite{Johannsen:2011dh}.


\vspace{0.1cm}
\noindent{\em \textbf{Acknowledgments}.}
  We thank Nico Yunes and Leo Stein for interesting discussions. We acknowledge the CENTRA-IST, in the Universidade T\'ecnica de Lisboa and the Universidade Federal do Par\'a for the kind hospitality during the period when this work was done.
   This work was partially supported by Coordena\c{c}\~ao de Aperfei\c{c}oamento de Pessoal de N\'{\i}vel Superior (CAPES), in Brazil, by Funda\c{c}\~ao para a Ciencia e a Tecnologia (FCT), in Portugal, by Conselho Nacional de Desenvolvimento Cient\'{\i}fico e Tecnol\'ogico (CNPq) and Funda\c{c}\~ao de Amparo \`a Pesquisa do Estado do Par\'a (FAPESPA), in Brazil, by the {\it DyBHo--256667} ERC Starting Grant and by FCT - Portugal through PTDC projects FIS/098025/2008, FIS/098032/2008, CTE-AST/098034/2008 and the grant SFRH/BD/44321/2008.  

\appendix
\section{Slowly rotating approximation}\label{app:derivation}
The modified field equations are obtained by varying the action~\eqref{action} with respect to the metric and to the scalar field. Varying the action~\eqref{action} with respect to the metric, neglecting the $S_{mat}$ term and the potential $V(\phi)$, we find 
\be
G_{ab}+\alpha_1\mathcal{H}_{ab}+\alpha_2\mathcal{I}_{ab}+\alpha_3\mathcal{J}_{ab}+\alpha_4\mathcal{K}_{ab}= T_{ab}^{\phi},
\label{fieldeq}
\ee
where $T^{\phi}_{ab}=\nabla_a\phi\nabla_b\phi-\frac{1}{2}g_{ab}\nabla_c\phi\nabla^c\phi$
and $\mathcal{H}_{ab}$, $\mathcal{I}_{ab}$, $\mathcal{J}_{ab}$, $\mathcal{K}_{ab}$ are explicitly given in Ref.~\cite{Yunes:2011we}.
Varying the action~\eqref{action} with respect to the scalar field $\phi$, we get
\be
-2\square\phi=\alpha_1 R^2 +\alpha_2 R_{ab}R^{ab} + \alpha_3 R_{abcd}R^{abcd}+\alpha_4 {R_{abcd}}^{*}R^{abcd}\nn\,.
\ee

We shall neglect terms of order $\alpha_i^2$ in the equation above. Since the Ricci scalar and the Ricci tensor are both zero in the background spacetime, the scalar field equation reduces to
\be
\square\phi=-\frac{1}{2}(\alpha_3 \tilde{R}_{abcd}\tilde{R}^{abcd} +\alpha_4\,^*\tilde{R}_{abcd}\tilde{R}^{abcd}),
\label{pertphieq}
\ee
where the tilde stands for background quantities. We note here that, when $a=0$, we recover the scalar field for spherically symmetric Gauss-Bonnet BHs \cite{Campbell:1991kz}, since at this order the Gauss-Bonnet term is just the Kretschmann invariant and there is no correction from the Chern-Simons term \cite{Yunes:2009hc}. On the other hand, for $\alpha_3=0$ we recover the scalar field for slowly rotating Chern-Simons BHs. Also, there is no correction of order $\alpha_3 a$, since the Kretschmann invariant has only corrections in even powers of $a$. Therefore, we can write
\be
\phi=\phi^{GB,CS}+\alpha_3 a^2 \phi^c(t,\theta),
\label{phiexp}
\ee
where $\phi^{GB,CS}$ is the scalar field for spherically symmetric Gauss-Bonnet BHs plus the correction of slowly rotating Chern-Simons BHs, both assuming small coupling constants. Substituting Eq. \eqref{phiexp} in Eq. \eqref{pertphieq}, we find that the only solution for $\phi^c$ which is regular at the horizon and goes to zero in the limit $\frac{r}{\mathcal{M}}\gg 1$, is given by the corresponding term in Eq. \eqref{scalar}.

Considering corrections up to $\alpha_i^2$, the modified Einstein's equations read
\bea
G_{ab} &+& 8 \alpha_3 \tilde{R}_{abcd}\tilde{\nabla}^c\tilde{\nabla}^d\phi +8 \alpha_4\,^*\tilde{R}_{(a}\,^c\,_{b)}\,^d\tilde{\nabla}_d\tilde{\nabla}_c\phi\nonumber\\
&&= \frac{1}{2}\l(2\tilde{\nabla}_a\phi\tilde{\nabla}_b\phi-\tilde{g}_{ab}\tilde{\nabla}_c\phi\tilde{\nabla}^c\phi\r),
\eea
in which the scalar field $\phi$ is given by Eq. \eqref{scalar}. We note here that the lowest dynamical corrections to the metric are given by second order terms in $\alpha_3$ and $\alpha_4$. 
Therefore, we can write
\be
g_{ab}=g_{ab}^{GB,CS}+\alpha_3^2 a g_{ab}^c\,\,,
\ee
where $g_{ab}^{GB,CS}$ is the metrics for the spherically symmetric Gauss-Bonnet BH plus the correction for slowly rotating Chern-Simons BH, both assuming small coupling constants. In the slowly rotating regime, the only non-vanishing term in $g_{ab}^c$ is $g_{t\varphi}^c$ \cite{Pani:2009wy}. With the ansatz $g_{t\varphi}^c=-H(r)\sin^2\theta$, we find that the only solution for $H(r)$ that goes to zero in the regime $\frac{r}{\mathcal{M}}\gg 1$, is given by the corresponding term in  Eq. \eqref{crosscomponent}.

\bibliographystyle{h-physrev4}
\bibliography{slowlyrotBHs}
\end{document}